\documentclass[twocolumn,superscriptaddress,amssymb,amsmath,aps,prb,showpacs]{revtex4-1}  

\usepackage{graphicx}
\usepackage{graphics}

\usepackage{color}

\begin{document}

\title{Correlation effect in Sr$_{1-x}$La$_x$RuO$_3$ studied by soft x-ray photoemission spectroscopy}

\author{Ikuto~Kawasaki}
\email{kawasaki@sci.u-hyogo.ac.jp}
\affiliation{Graduate School of Material Science, University of Hyogo, Kamigori, Hyogo 678-1297, Japan}

\author{Yumi~Sakon}
\affiliation{Faculty of Science, Ibaraki University, Mito 310-8512, Japan}
\author{Shin-ichi~Fujimori}
\affiliation{Condensed Matter Science Division, Japan Atomic Energy Agency, Sayo, Hyogo 679-5148, Japan}
\author{Hiroshi~Yamagami}
\affiliation{Condensed Matter Science Division, Japan Atomic Energy Agency, Sayo, Hyogo 679-5148, Japan}
\affiliation{Department of Physics, Faculty of Science, Kyoto Sangyo University, Kyoto 603-8555, Japan}
\author{Kenichi~Tenya}
\affiliation{Faculty of Education, Shinshu University, Nagano 390-8621, Japan} 
\author{Makoto~Yokoyama}
\affiliation{Faculty of Science, Ibaraki University, Mito 310-8512, Japan}

\date{\today}%

\begin{abstract}

To clarify how the electronic state of Sr$_{1-x}$La$_x$RuO$_3$ evolves with La doping, we conducted photoemission (PES) experiments using soft x-rays. The spectral shape of the Ru 4$d$ derived peak near the Fermi level changes significantly with increasing $x$. This variation indicates that a spectral weight transfer  from the coherent to incoherent component occurs due to an enhancement of the electron correlation effect. Resonant PES experiments at the La 3$d_{5/2}$ edge have  confirmed that there is no significant contribution of the La 5$d$ state in the energy range where the spectral weight transfer is observed. Using the dependence of the photoelectron  mean free path on the photon energy, we subtracted the surface components from the PES spectra and  confirmed that the enhancement of the electron correlation effect with La doping is an intrinsic bulk phenomenon. On the other hand,  a large portion of the coherent component remains at the Fermi level up to $x$ = 0.5, reflecting that the Ru 4$d$ state still has itinerant characteristics. Moreover, we found that the PES spectra hardly depend on the temperature and do not exhibit a discernible change with magnetic ordering, suggesting that the temperature variation of the exchange splitting does not follow the prediction of the  Stoner theory. The presently obtained experimental results indicate that the electron correlation effect plays an important role in Sr$_{1-x}$La$_x$RuO$_3$ and that
the Ru 4$d$ electrons possess both local and itinerant characteristics.

\end{abstract}


\maketitle

\section{INTRODUCTION}

SrRuO$_3$ is one of the rare  4$d$ transition-metal oxides that shows a ferromagnetic order and has attracted considerable interest due to its fascinating electronic and magnetic properties.
This compound has a pseudocubic perovskite structure (a GdFeO$_3$ type orthorhombic structure). Its Curie temperature $T_\mathrm{C}$ and ordered ferromagnetic moment are 160 K and approximately 1$\mu_\mathrm{B}$/f.u., respectively.\cite{calla,kanba} 
The  fundamental and important question concerning this compound is how the electron correlation effect affects its electronic properties. At first glance,  because Ru 4$d$ electrons are generally expected to have a delocalized nature in the crystal, it is natural to consider that the electron correlation effect is not significant and that the ferromagnetic order  can be described well in terms of the Stoner theory in an itinerant electron picture. However, this compound shows so-called bad metal behavior, and its resistivity exceeds the Ioffe-Regel limit at high temperatures, indicating a breakdown of the quasi-particle description.\cite{allen,gunn}  Optical spectroscopy has shown that the charge dynamics deviate from  Fermi liquid behavior.\cite{kostic,dodge} Moreover, the Rhodes-Wohlfarth ratio $p_c$/$p_s$, where $p_c$ and $p_s$ represent the paramagnetic dipole moment estimated from the Curie constant and the ordered  moment at zero temperature, respectively, is approximately 1.3.\cite{fukunaga}  This value indicates that SrRuO$_3$  is close to the localized-moment limit. These results imply that the  correlation effect of the Ru 4$d$ electrons plays an important role and that the electronic state is not conventional.

Photoemission (PES) spectroscopy provides a direct probe into the electronic band structure. A number of PES experiments have been carried out to clarify the electronic state of SrRuO$_3$. Earlier PES studies reported that although the overall structure of the valence spectra consisting of the Ru 4$d$ and O 2$p$ states was roughly reproduced by  calculations based on the density functional theory (DFT), the Ru 4$d$ derived peak near the Fermi level ($E_\mathrm{F}$) was strongly suppressed compared to that derived from DFT calculations.\cite{fujio_ph,okamoto_ph,park_ph}   However, subsequent PES studies have revealed, by using bulk sensitive soft x-rays and/or $in$ $situ$ grown thin film samples, that PES spectra are very sensitive to surface conditions, and the valence spectra of the earlier PES studies were largely influenced by the extrinsic surface electronic states.\cite{siemon_ph,kim_ph,horiba_ph,takizawa_ph,kumigashira_ph,toyota_ph,grebin_ph}  These subsequent PES studies demonstrated that their valence spectra exhibit a pronounced peak originating from the Ru 4$d$ state just below $E_\mathrm{F}$ as predicted by DFT calculations, showing that  DFT calculations based on an itinerant electron picture can be a good starting point to describe the electronic state of SrRuO$_3$. However, following experimental observations in PES investigations suggest that the Ru 4$d$ electrons also have localized characteristics. The PES spectra show an incoherent feature, which is not reproduced in DFT calculations. This incoherent feature is possibly originating from the electron correlation effect and has been reproduced in a recent calculation based on the combination of the DFT and the dynamical mean-field theory (DFT+DMFT), which can incorporate the electron correlation effect beyond the conventional DFT.\cite{kim_dmft} In addition,  recent angle-resolved PES and optical conductivity experiments  have shown that the temperature dependence of ferromagnetic exchange splitting is markedly weaker than that predicted in the Stoner theory.\cite{shai,jeong} It has been suggested in these studies that  exchange splitting persists above $T_\mathrm{C}$ but that the long-range ferromagnetic order is destroyed by the spatial and temporal fluctuations of the exchange splitting in a local region above $T_\mathrm{C}$. This fluctuating exchange splitting in a local region above $T_\mathrm{C}$ is reminiscent of localized magnetic moments. Therefore, it is considered that the Ru 4$d$ electrons in SrRuO$_3$ have both local and itinerant characteristics.

In fact, the Ru 4$d$ electrons in SrRuO$_3$ have an instability toward electron localization due to the  correlation effect. 
Substituting Mn for Ru suppresses the ferromagnetic order and induces an antiferromagnetic insulating phase above $\sim$35\% of Mn concentration.\cite{cao_Mn,sahu_Mn,yokoyama_Mn,kolesnik_Mn} Similar insulating phases have been found for several other Ru site substituted systems.\cite{kimlee,willgilli} 
The substitution of La for Sr also reduces the ferromagnetic order.\cite{bouchard,nakatsu} We have recently reported that the ferromagnetic order in Sr$_{1-x}$La$_x$RuO$_3$ changes into a cluster-glass state for $x$ $\geq$ 0.3, whereas the metallic character is retained  up to at least $x$ = 0.5.\cite{kawasaki_macro}  The ac susceptibility depends markedly on the frequency near the cluster-glass ordering temperature $T^*$,  and the frequency dependence of $T^*$  follows the Vogel-Fulcher law.  Furthermore, $\mu$SR investigations for $x$ $\geq$ 0.3 demonstrated that magnetic clusters start developing well above $T^*$ and that the magnetic ordering process of the cluster-glass state is strikingly different from that expected for  a conventional long-range ferromagnetic order.\cite{kawasaki_mu} Note that the emergence of a cluster-glass state is an indication of the localized character of the Ru 4$d$ states  because it reflects the presence of  spatial inhomogeneities in the magnetism.   We have also revealed that the electronic specific-heat coefficient increases significantly with La doping, indicating that  La doping may enhance the effective mass of the conduction electrons due to the electron correlation effect.\cite{kawasaki_macro} In this study, we performed PES measurements using soft x-rays to study how the electronic state of Sr$_{1-x}$La$_x$RuO$_3$ evolves with La doping  and to observe the tendencies of the electron localization indicated by the emergence of the cluster-glass state.

\section{EXPERIMENTAL DETAILS}

Polycrystalline samples of Sr$_{1-x}$La$_x$RuO$_3$ for $x$ $\leq$ 0.5 were prepared by the conventional solid-state reaction method as reported previously.\cite{kawasaki_macro} PES experiments were carried out at the soft x-ray beamline BL23SU in  SPring-8.\cite{BL23SU} The base pressure in the sample chamber was maintained below  2$\times$10$^{-8}$ Pa during the measurements.  To obtain clean surfaces, the samples were fractured $in$ $situ$ just before the measurements under  ultrahigh vacuum conditions. The PES spectra were measured using a VG-SCIENTA SES2002 analyzer. The energy resolution was approximately 120 meV at a photon energy of $h\nu$ = 700 eV and increased up to 300 meV at $h\nu$ = 1100 eV. The position of $E_\mathrm{F}$ was calibrated by measuring the Fermi edge of an evaporated gold film. The sample temperature was controlled by a He-flow cryostat, and the PES spectra were collected over a temperature range from 25 to 200 K. Measurements of the x-ray absorption spectrum (XAS) were performed in the total-electron-yield mode.

\section{RESULTS}

\begin{figure}[t]
\begin{center}
\includegraphics[keepaspectratio, width=7cm,bb = 0 0 240 310,clip]{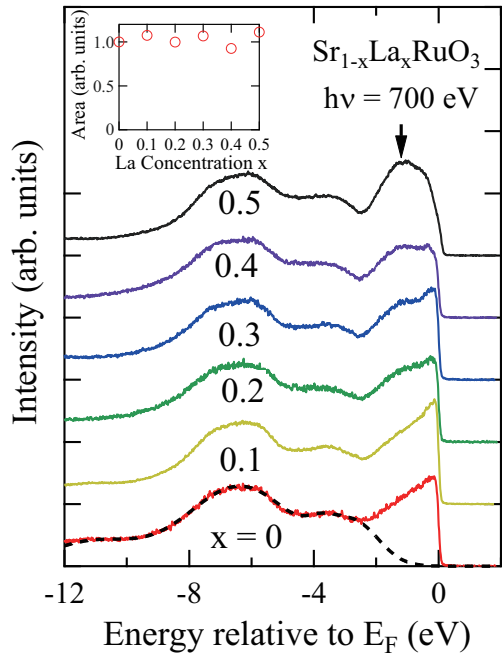}
\end{center}
\caption{(Color online) Photoemission spectra of Sr$_{1-x}$La$_x$RuO$_3$ measured at $h\nu$ = 700 eV. The spectra are normalized by the spectral area from -2.5 to -10 eV, which roughly corresponds to the total intensity of the O 2$p$ states. The position of the incoherent component is indicated by the arrow. The black broken line represents the fit of the spectral weight of the O 2$p$ states using  multiple Gaussians. The inset shows the $x$ variation of the  spectral area from 0.5 to -2.5 eV, which approximately corresponds to the total number of Ru 4$d$ electrons contributing to the peak just below $E_\mathrm{F}$. }
\label{f1}
\end{figure}

Figure 1 shows the PES spectra of Sr$_{1-x}$La$_x$RuO$_3$. The sample temperature was maintained at 25 K during  data collection. At this temperature, the samples  with $x$ $\leq$ 0.2 were in ferromagnetic ordered states, and samples with $x$ $\geq$ 0.3, which exhibit cluster-glass states at low temperatures, were in paramagnetic states.\cite{kawasaki_macro,kawasaki_mu} However, it should be mentioned here that the valence spectra hardy depend on the temperature in the temperature and $x$ range investigated here, as shown later in this section. The incident photon energy was 700 eV.  At this photon energy,  the Ru 4$d$, O 2$p$, and La 5$d$ states have relatively large PES cross-sections, whereas the contributions from other valence electrons, such as Ru 5$s$, Sr 5$s$,  and La 6$s$, are essentially negligible because their contributions are approximately two orders of magnitude smaller than that of  Ru 4$d$.\cite{lindau}       
The spectrum for $x$ = 0 is very similar to those of previous PES studies using soft x-rays.\cite{horiba_ph,takizawa_ph,kumigashira_ph,toyota_ph}
The peak just below $E_\mathrm{F}$ extending to approximately $-2$ eV originates from the Ru 4$d$ states, and the intensity distributed from  $-2$ to $-10$ eV is mainly attributed to the O 2$p$ states.\cite{grebin_ph} The line shape of the intensity of the O 2$p$ states does not exhibit any noticeable change  with La doping.  We normalize these spectra by the spectral area from -2.5 to -10 eV, which roughly corresponds to the total intensity of the O 2$p$ states. Although the La 5$d$ states, whose PES cross-sections are comparable that of O 2$p$ at $h\nu$ = 700 eV,\cite{lindau} also contribute to this energy range with increasing $x$ as shown later in this section, the La 5$d$ contribution compared to the total intensity of the O 2$p$ states is less than few percent and essentially negligible by the following reason. The total number of O 2$p$ electrons is roughly 18/f.u. assuming that the O 2$p$ bands are nearly fully occupied as predicted by DFT calculations,\cite{maiti} whereas the total number of La 5$d$ electrons is considered to be less than 0.5 at most for  $x$ $\leq$ 0.5. The shape of the Ru 4$d$ derived peak shows remarkable change; the intensity just below $E_\mathrm{F}$ is suppressed with increasing $x$, and the intensity develops near  $-1.2$ eV (indicated by the arrow in Fig. 1).  It has generally been recognized in previous PES studies that the peak just below $E_\mathrm{F}$ and the broad spectral weight near $-1.2$ eV in SrRuO$_3$ correspond to the coherent and incoherent components of the Ru 4$d$ state, respectively.\cite{fujio_ph,okamoto_ph,park_ph,siemon_ph,kim_ph,horiba_ph,takizawa_ph,grebin_ph}  The presence of the incoherent component near $-1.2$ eV is supported by a recent DFT+DMFT calculation for SrRuO$_3$,\cite{kim_dmft} indicating that it is originating from the electron correlation effect.  In this context, the observed spectral change suggests that a spectral weight transfer of the Ru 4$d$ electrons from the coherent to incoherent components occurs upon doping La.  We also display the $x$ variation of the spectral area from 0.5 to -2.5 eV, which roughly corresponds to  the total number of Ru 4$d$ electrons contributing to the peak just below $E_\mathrm{F}$, in the inset of Fig. 1. Within the experimental accuracy,  the spectral area from 0.5 to -2.5 eV seems unchanged, suggesting an absence of a large change in the valence state of Ru with La doping. However, we cannot discuss the detailed behavior of the valence state evolution of Ru by La doping due to the presence of the data scatter of about 10\%.

\begin{figure}[b] 
\begin{center}
\includegraphics[keepaspectratio, width=7cm,bb = 0 0 240 480,clip]{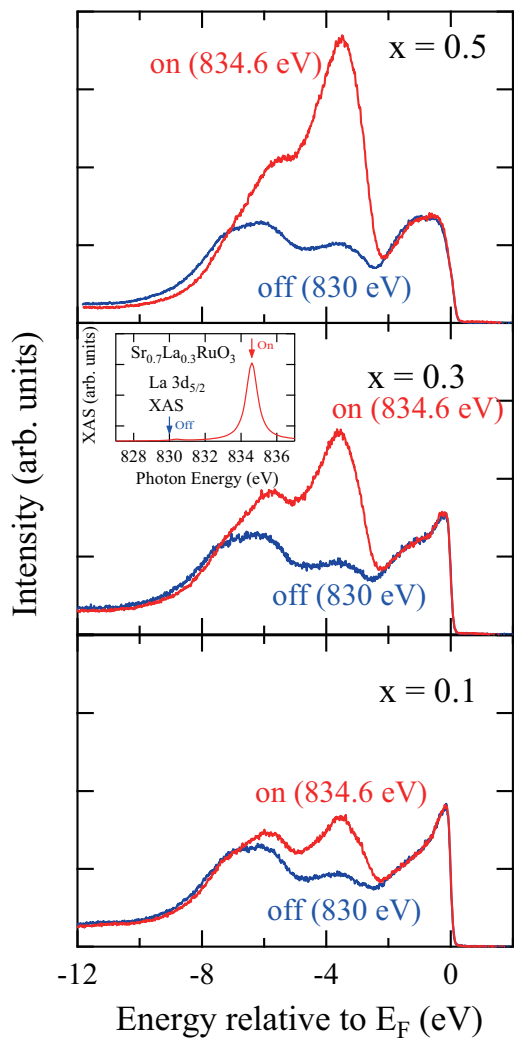}
\end{center}
\caption{(Color online)  La 3$d_{5/2}$ on-resonance and off-resonance photoemission spectra of Sr$_{1-x}$La$_x$RuO$_3$. These spectra were normalized by the intensity around $E_\mathrm{F}$.  The inset shows the  La 3$d_{5/2}$ x-ray absorption spectrum for $x$ = 0.3.  }
\label{f1}
\end{figure}

Since the doped La atoms may provide additional La 5$d$ electrons to the valence bands, one may consider that the change of the  peak near $E_\mathrm{F}$ is caused not by the spectral weight transfer of the Ru 4$d$ state from the coherent to incoherent components but by the development of the La 5$d$ component.  To check this possibility, we performed La 3$d$-edge resonance PES experiments to identify the La 5$d$ contributions in the valence band spectra.\cite{sekiyama}  In the La 3$d$-edge resonance PES, the following two processes occur simultaneously:
\begin{eqnarray}
3d^{10}4f^{0}5d^n + h\nu   &\rightarrow& 3d^{10}4f^{0}5d^{n-1} + \mathrm{electron},\\
3d^{10}4f^{0}5d^n + h\nu   &\rightarrow& 3d^{9}4f^{1}5d^n \notag \\ 
&\rightarrow& 3d^{10}4f^{0}5d^{n-1} + \mathrm{electron}.  
\end{eqnarray}  
These are known as the (1) direct  and (2) Auger  processes. Because these two processes have the same initial and final states, the La 5$d$ spectral weight is expected to be resonantly enhanced when the photon energy corresponds to the La 3$d$-edge. The inset of Fig. 2 shows the La 3$d_{5/2}$ XAS spectrum for $x$ = 0.3. A peak is clearly seen at a photon energy of $h\nu$ = 834.6 eV. The incident photon energy was chosen to this peak energy for the on-resonance measurements and $h\nu$ = 830 eV for the off-resonance measurements. The on- and off-resonance spectra measured at 25 K for each sample are displayed in Fig. 2. It is clear that the intensity of the peak at around $-3.5$ eV is markedly enhanced in the on-resonance spectra.  We normalized these  on- and off-resonance spectra by the intensity around $E_\mathrm{F}$ and found that the on- and off-resonance spectra overlap almost perfectly around $E_\mathrm{F}$. This reflects an absence of the La 5$d$ component in this energy range, where the Ru 4$d$ component is dominant. Here, we note that this normalization method is almost equivalent to that by the incident photon intensity since the difference in the PES cross-section of Ru 4$d$ between the  on- and off-resonance spectra is only about 1\%.\cite{lindau} 
Because the La 5$d$ contribution is limited to the energy range between $-2$ and $-7.5$ eV for all the samples presently investigated, it is considered that the La 5$d$ states mainly hybridize with the O 2$p$ states and do not influence the change of the spectra near $E_\mathrm{F}$. It should be also noted here that the presence of La 5$d$ component contradicts with the simple ionic picture, since the La atom is generally considered to be trivalent in this picture, and thus, the La 5$d$ state should be empty.\cite{bouchard,nakatsu} The presence of La 5$d$ state is consistent with our recent magnetic susceptibility measurements, which showed that the $x$ variation of  the effective moment cannot be explained in terms of the simple ionic picture.\cite{kawasaki_macro}  Since Sr$^{2+}$ is substituted by La$^{3+}$, the same fraction of Ru$^{4+}$ is expected to be substituted by Ru$^{3+}$ through electron transfer in the simple ionic picture. The Ru$^{4+}$ and Ru$^{3+}$ ions have effective moments of 2.83 and 1.73$\mu_\mathrm{B}$, respectively,\cite{bouchard} and therefore, the effective moment is expected to  decrease monotonously with increasing $x$ in the simple ionic picture. However, the experimentally obtained effective moment  estimated from the Curie constant does not shows a monotonous decrease with $x$. Both the present La 3$d$-edge resonance PES and our recent magnetic susceptibility measurements imply that the Ru 4$d$ state in Sr$_{1-x}$La$_x$RuO$_3$ cannot be understood in terms of the simple ionic configurations of Ru$^{4+}$ and Ru$^{3+}$ possibly because of its delocalized nature.

\begin{figure}[t]
\begin{center}
\includegraphics[keepaspectratio, width=7cm,bb = 0 0 350 630,clip]{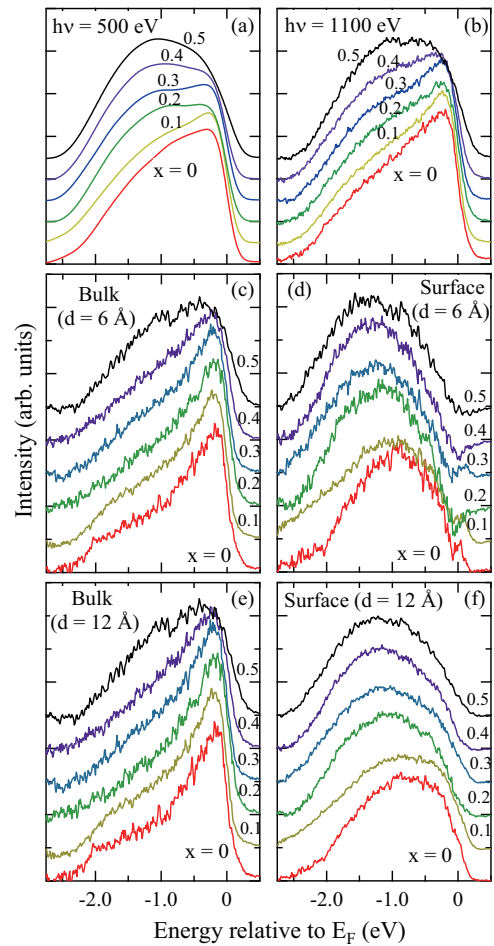}
\end{center}
\caption{(Color online) The Ru 4$d$ components extracted from photoemission spectra  measured at (a) $h\nu$ = 500 and (b) 1100 eV, which are obtained by subtracting the spectral weight at higher binding energies dominated by the O 2$p$ states. These spectra are normalized by the area of each spectrum. The results shown in (c)-(f) are the bulk and surface contributions estimated by using Eq. (3) for $d$ = 6 and 12 $\mathrm{\AA}$, where the parameter $d$  represents the the thickness of the surface layer.  }
\label{f1}
\end{figure}

Since the PES experiments were carried out in the soft x-ray range, the PES spectra mainly reflect the bulk electronic states. However, the influence from the surface region  of Sr$_{1-x}$La$_x$RuO$_3$ may not be negligible.  In fact, the electronic state in the surface region of SrRuO$_3$ seems to be markedly different from that of the bulk state, since the PES spectra obtained for polycrystal samples  with scraped or fractured surfaces largely depend on the photon energy, namely, on the probing depth.\cite{fujio_ph,okamoto_ph,horiba_ph} Here, we subtract the surface components from the PES spectra according to a procedure described in Ref. 14  to investigate the intrinsic bulk electronic properties.  
The PES spectra can be decomposed into surface and bulk components using the following equation:
\begin{eqnarray}
I(E) = I_\mathrm{surface}(E) (1-e^{-d/\lambda}) +  I_\mathrm{bulk}(E)e^{-d/\lambda},
\end{eqnarray} 
where $I_\mathrm{surface}$ and $I_\mathrm{bulk}$ represent the intensities of the surface and bulk components, respectively. $d$ is the thickness of the surface layer, and $\lambda$  is the photoelectron mean free path. Since $\lambda$ depends on the  photon energy, we can estimate the surface and bulk components from two spectra measured at different photon energies. In this analysis, the O 2$p$ spectral weights at higher binding energies  are fitted by multiple Gaussians as shown in Fig. 1, and then we subtract them from the PES spectra to focus on the behavior of the Ru 4$d$ derived peak.   The Ru 4$d$ derived peaks in the PES spectra measured with  $h\nu$ = 500 and 1100 eV at 25 K are shown in Figs. 3(a) and 3(b), and these spectra are normalized by the area of each spectrum. The spectra measured at $h\nu$ = 500 eV are broadened using a Gaussian to decrease the energy resolution to that of $h\nu$ = 1100 eV so that we can compare the spectra.  We estimated  $\lambda$ to be approximately 10 and 19 $\mathrm{\AA}$ for  $h\nu$ = 500 and 1100 eV, respectively, using semiempirical expressions given in Ref. 35. The parameter $d$ is assumed to be same order in magnitude with lattice constants,\cite{nakatsu} and we estimated the surface and bulk components by tentatively using $d$ values of  6 and 12 $\mathrm{\AA}$ [Figs. 3(c)-3(f)]. Figures 3(c) and 3(e) clearly show that the spectral shape of the bulk component does not exhibit any noticeable change with the doubling of $d$, suggesting that the bulk spectra estimated in this procedure are robust against  estimation errors in $d$. Therefore, the estimated bulk spectra would reflect the intrinsic bulk electronic states, even though the accurate value of $d$ is unknown. Although the incoherent component ($\sim$1.2 eV)  is stronger in the surface spectra, one can see the spectral weight transfer from the coherent to incoherent component upon La doping in the bulk spectra.  The strong enhancement of the incoherent component in the surface spectra can naturally be understood by considering the fact that the surface Ru 4$d$ state is expected to be less hybridized than the bulk Ru 4$d$ state. This is because the surface atoms have fewer nearest neighbors. Hence, this enhancement  further supports that the incoherent component  at $\sim$1.2 eV originates from the electron correlation effect.

\begin{figure}[t]
\begin{center}
\includegraphics[keepaspectratio, width=7cm,bb = 0 0 250 630,clip]{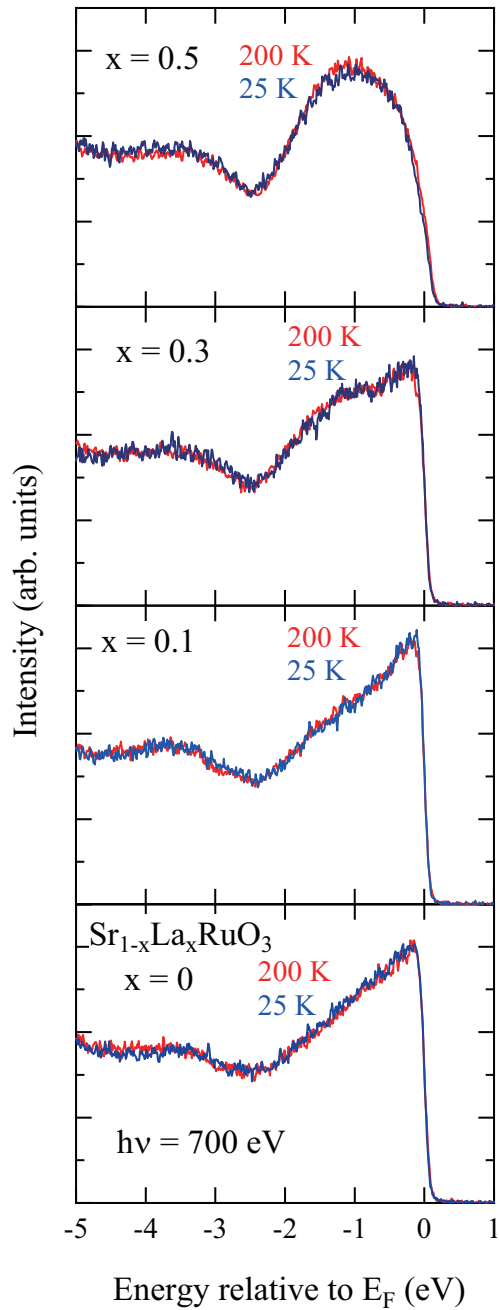}
\end{center}
\caption{(Color online) Temperature dependence of the photoemission spectra for Sr$_{1-x}$La$_x$RuO$_3$ measured at $h\nu$ = 700 eV. These spectra are normalized by the area of each spectrum from 0.5 to $-5$ eV.   }
\label{f1}
\end{figure}

We measured the temperature dependence of the PES spectra to study how magnetic ordering affects the valence electronic state of Sr$_{1-x}$La$_x$RuO$_3$. Figure 4 shows the PES spectra for each sample measured at 25 and 200 K.  These spectra are normalized by the area of each spectrum. The ferromagnetic ordering temperatures for $x$ = 0 ($T_\mathrm{C}$ = 160 K) and 0.1 ($T_\mathrm{C}$ = 120 K) are between 25 and 200 K and well separated from these temperatures. Although the cluster-glass ordering temperatures for $x$ $\geq$ 0.3 are lower than 25 K,  the magnetic clusters already start developing at this temperature according to our recent $\mu$SR study.\cite{kawasaki_mu}  The PES spectra exhibit no discernible dependence on the temperate over the investigated  $x$ range.

\section{DISCUSSION}
The present study reveals that the substitution of La for Sr generates a clear weight transfer of the Ru 4$d$ state from the coherent to incoherent component.  Since the incoherent feature is reproduced by a recent DFT+DMFT  calculation for SrRuO$_3$, which can take account of the man- body effect beyond the conventional DFT calculations, \cite{kim_dmft} the development of the incoherent component seems to be caused by the enhancement of electron correlation by La doping. 
The enhancement of electron correlation is consistent with the emergence of cluster-glass states for $x$ $\geq$ 0.3 because both are indicative of the  localized character of the Ru 4$d$ electrons. On the other hand, it is important to note that the large PES intensity at $E_\mathrm{F}$ remains up to $x$ = 0.5 despite the development of the incoherent component. This  suggests that the Ru 4$d$ state still involves itinerant characteristics and is consistent with the metallic character of Sr$_{1-x}$La$_x$RuO$_3$ revealed by resistivity and specific-heat measurements.\cite{kawasaki_macro}

Next, we  argue the relationship between this variation in the PES spectra with La doping and the thermodynamic quantities. We previously demonstrated that low temperature magnetization is markedly suppressed with La doping.\cite{kawasaki_macro} This indicates that exchange splitting at low temperatures is reduced with increasing $x$. A DFT calculation for SrRuO$_3$ predicts that the reduction of the exchange splitting accompanies the enhancement of the density of states near $E_\mathrm{F}$ and therefore makes the  Ru 4$d$ derived peak more pronounced.\cite{takizawa_ph} However, the present PES results show an opposite behavior with increasing $x$. Hence, it is considered that the effect of the reduction in exchange splitting on the PES spectra is overwhelmed by the suppression of the PES intensities near  $E_\mathrm{F}$ due to the renormalization factor $Z$ = $(1- \partial \mathrm{Re}\Sigma /\partial \omega)^{-1}$, where $\Sigma$ is the self-energy originating from the electron correlation effect. We previously reported that the electronic specific-heat coefficient $\gamma$ exhibits a monotonic increase with increasing $x$ up to $x$ = 0.3, which is accompanied by a significant suppression in the magnetic ordering temperature and the ordered moment.\cite{kawasaki_macro} The increase rate in  $\gamma$ with La doping becomes small for $x$ $\geq$ 0.3, where the low-temperature ordered moment is well suppressed. Combined with the features of the PES spectra, we consider that the enhancement in $\gamma$ up to $x$ = 0.3 is caused by a combination of the suppression of exchange splitting and  the development of the electron correlation effect. In contrast, the weak increase in $\gamma$ for  $x$ $\geq$ 0.3 likely originates solely from the latter effect.

Here, we discuss the origin of the enhancement of electron correlation and the  suppression of the ferromagnetic order with La doping. According to Ref. 28, the La doping increases the lattice constants and also enhances the GdFeO$_3$ type distortion by tilting the RuO$_6$ octahedra. The former change in the crystal structure would enhance the density of states at $E_\mathrm{F}$ by weakening the hybridization effect. In contrast, the latter change may diminish the density of states at $E_\mathrm{F}$ by lifting the degeneracy of the Ru 4$d$ derived bands near $E_\mathrm{F}$, which have $t_{2g}$ symmetry.\cite{dang} At first glance, the enhancement of electron correlation with La doping revealed by this study seems to suggest that the evolution of the electronic state caused by the former change is dominant in Sr$_{1-x}$La$_x$RuO$_3$. However, in this explanation, the origin of the suppression of ferromagnetic would be puzzling. This is because the presence of the ferromagnetic order is determined by the value of the product of the density of states at $E_\mathrm{F}$ and an interaction constant in the Stoner theory.\cite{stoner} Therefore, in this sense, the suppression of the ferromagnetic order can be considered as an indication of the suppression of the density of states at $E_\mathrm{F}$ and the broadening of the Ru 4$d$ bands near $E_\mathrm{F}$. This is clearly inconsistent with the above explanation for the origin of the enhancement of electron correlation. Further experimental and theoretical investigations are clearly needed to understand this puzzling coexistence of the suppression of the ferromagnetic order and the enhancement of  electron correlation with La doping.

It is remarkable that the PES spectra exhibit no significant temperature variations in contrast to their clear $x$ dependence.  X-ray magnetic circular dichroism  spectroscopy studies have revealed that orbital moments are almost quenched in the ferromagnetic state of SrRuO$_3$.\cite{agrestini,okamoto} Therefore, the ferromagnetic order originates from spin moments, and a considerable amount of exchange splitting  between the majority and minority spin bands is indispensable to reproduce the ordered moment of approximately 1$\mu_\mathrm{B}$.\cite{calla,kanba}   In fact, the energy scale of the exchange splitting for  $x$ = 0 is  estimated to be approximately 0.5 eV by DFT calculations.\cite{maiti} This value is  much larger than the energy resolution of present PES study (120 meV at $h\nu$ = 700 eV). Therefore, the temperature dependence of the PES spectra should be measurable, if the conventional Stoner theory is applicable to the ferromagnetic order in this compound. This is because PES spectra are the simple sum of the majority and minority spin components, and the exchange splitting between these components should decrease with increasing temperature and disappear above $T_\mathrm{C}$  in the Stoner theory. 
Actually, Stoner-like temperature dependence has been reported in PES experiments for ferromagnetic Gd metal,\cite{Gd_kim} although subsequent detailed spin-resolved PES experiments revealed that the temperature dependence is much more complex than the prediction of the Stoner theory.\cite{Gd_maiti}  Our PES data are consistent with recent angle-resolved PES and optical spectroscopy studies for $x$ = 0,\cite{shai,jeong}  which indicate that  exchange splitting does not disappear at $T_\mathrm{C}$ and remains finite well above $T_\mathrm{C}$. In these studies,  it has been proposed that such a residual exchange splitting above $T_\mathrm{C}$  can be explained based on fluctuating local band theory.\cite{koreman} In this theory, the band structure in a local region still has  exchange splitting above $T_\mathrm{C}$; however, the long-range magnetic order is destroyed by spatial and  temporal fluctuations. The recent DFT+DMFT calculation  also reproduced residual  exchange splitting above $T_\mathrm{C}$.\cite{kim_dmft} 

Note that the PES spectra of Sr$_{1-x}$La$_x$RuO$_3$ do not show a discernible temperature dependence for the entire investigated $x$ range, whereas the ferromagnetic transition temperature varies widely with  $x$.  $T_\mathrm{C}$ and the ordered moment in the ferromagnetic state  for $x$ = 0.1 are approximately 120 K and 0.75$\mu_\mathrm{B}$, respectively, which are approximately 75\% of those for SrRuO$_3$.\cite{kawasaki_macro} The present PES investigations imply that the exchange splitting for $x$ = 0.1 also remains nearly unchanged up to 200 K, which is well above  $T_\mathrm{C}$. In this context, it is natural to speculate that exchange splitting  for $x$ $\geq$ 0.3 also remains finite above the cluster-glass ordering temperatures, and it is closely related to the origin of the development of magnetic clusters well above the cluster-glass ordering temperatures observed in the $\mu$SR study.\cite{kawasaki_mu} Even though the absence of temperate dependence in the PES spectra for $x$ $\geq$ 0.3 seem to be consistent with this speculation, this is not conclusive because the low-temperature magnetization for $x$ $\geq$ 0.3 is significantly smaller than that for SrRuO$_3$, and it may yield too small exchanging splitting to be detected in the present PES experiment.\cite{kawasaki_macro}

\section{CONCLUSION}

The electronic states of Sr$_{1-x}$La$_x$RuO$_3$ were studied by PES experiments using soft x-rays. The Ru 4$d$ derived peak near $E_\mathrm{F}$ is significantly influenced by the substitution of La for Sr; doping La into SrRuO$_3$ reduces the PES intensity just below $E_\mathrm{F}$,  accompanying the evolution of an incoherent component near  $-1.2$ eV. The resonant PES experiments at the La 3$d_{5/2}$ edge confirmed that the change in the  spectral shape of the Ru 4$d$ derived peak does not originate  from  the additional contribution of the La 5$d$ electrons. We decomposed the PES spectra into the bulk and surface components using the dependence of the photoelectron  mean-free path on the photon energy and found that  the spectral weight transfer of the Ru 4$d$ state from the coherent to incoherent components upon La doping occurs in the bulk component. The development of the incoherent component is considered to reflect the enhancement of the electron correlation effect with La doping. This is  consistent with the emergence of cluster-glass states for $x$ $\geq$ 0.3, which indicates the localized nature of the Ru 4$d$ electrons. On the other hand,  a large portion of the PES intensity  at $E_\mathrm{F}$ still remains even for $x$ = 0.5, showing that the Ru 4$d$ state still involves itinerant characteristics. The temperature dependence of the PES spectra is negligibly small and does not follow the prediction of the Stoner theory. These results suggest that the electronic and magnetic properties of Sr$_{1-x}$La$_x$RuO$_3$ are influenced by the electron correlation effect and that the  Ru 4$d$ electrons have both local and itinerant natures.

\acknowledgments

The PES experiments were carried out at the JAEA beamline BL23SU in SPring-8 (Proposal No. 2015B3884). This work was performed under the Shared Use Program of the JAEA Facilities (Proposal No. 2015B-E25) with the approval of the  Nanotechnology Platform project supported by the Ministry of Education, Culture, Sports, Science, and Technology (Proposal No. A-15-AE-0042).

\end{document}